\documentclass[default]{aastex631}

\begin{document}

\title{The JWST Early Release Science Program for Direct Observations of Exoplanetary Systems: Recommendations for Best Practices for Data Collection in Cycle 2 and Beyond
}

\correspondingauthor{Sasha Hinkley}
\email{S.Hinkley@exeter.ac.uk}

\author[0000-0001-8074-2562]{Sasha Hinkley}\affiliation{University of Exeter, Astrophysics Group, Physics Building, Stocker Road, Exeter, EX4 4QL, UK.}
\author{Beth Biller}\affiliation{SUPA, Institute for Astronomy, The University of Edinburgh, Royal Observatory, Blackford Hill, Edinburgh, EH9 3HJ, UK}
\author{Andrew Skemer}\affiliation{Department of Astronomy \& Astrophysics, University of California, Santa Cruz, CA 95064, USA}
\author[0000-0001-5365-4815]{Aarynn L.~Carter}\affiliation{Department of Astronomy \& Astrophysics, University of California, Santa Cruz, CA 95064, USA}
\author{Julien Girard}\affiliation{Space Telescope Science Institute, 3700 San Martin Drive, Baltimore, MD 21218, USA}
\author{Dean Hines}\affiliation{Space Telescope Science Institute, 3700 San Martin Drive, Baltimore, MD 21218, USA}
\author[0000-0003-2769-0438]{Jens Kammerer}\affiliation{Space Telescope Science Institute, 3700 San Martin Drive, Baltimore, MD 21218, USA}
\author[0000-0002-0834-6140]{Jarron Leisenring}\affiliation{Steward Observatory, University of Arizona, 933 N. Cherry Ave, Tucson, AZ 85721-0065 USA}
\author{William Balmer}\affiliation{Center for Astrophysical Sciences, The William H. Miller III Department of Physics and Astronomy, JHU, Baltimore, MD 21218, USA}
\author{Elodie Choquet}\affiliation{Aix Marseille Univ., CNRS, CNES, LAM, Marseille, France}
\author[0000-0001-6205-9233]{Maxwell A. Millar-Blanchaer}\affiliation{Department of Physics, University of California, Santa Barbara, Santa Barbara, CA, USA}
\author{Marshall Perrin}\affiliation{Space Telescope Science Institute, 3700 San Martin Drive, Baltimore, MD 21218, USA}
\author{Laurent Pueyo}\affiliation{Space Telescope Science Institute, 3700 San Martin Drive, Baltimore, MD 21218, USA}
\author{Jason Wang}\affiliation{Northwestern University}
\author[0000-0002-4479-8291]{Kimberly Ward-Duong}\affiliation{Department of Astronomy, Smith College, Northampton MA 01063, USA}\affiliation{Space Telescope Science Institute, 3700 San Martin Drive, Baltimore, MD 21218, USA}
\author{Anthony Boccaletti}\affiliation{LESIA, Observatoire de Paris, Universit\'e PSL, CNRS, Sorbonne Universit\'e, Universit\'e Paris Cit\'e, 5 place Jules Janssen, 92195 Meudon, France}
\author{Brittany Miles}\affiliation{Department of Astronomy \& Astrophysics, University of California, Santa Cruz, CA 95064, USA}
\author[0000-0001-8718-3732]{Polychronis Patapis}\affiliation{Institute for Particle Physics \& Astrophysics, ETH Zurich, 8092 Zurich, Switzerland}
\author[0000-0002-4388-6417]{Isabel Rebollido}\affiliation{Space Telescope Science Institute, 3700 San Martin Drive, Baltimore, MD 21218, USA}
\author[0000-0003-4203-9715]{Emily Rickman}\affiliation{European Space Agency (ESA), ESA Office, Space Telescope Science Institute, 3700 San Martin Drive, Baltimore, MD 21218, USA}
\author[0000-0001-9855-8261]{B.~Sargent}\affiliation{Space Telescope Science Institute, 3700 San Martin Drive, Baltimore, MD 21218, USA}\affiliation{Center for Astrophysical Sciences, The William H. Miller III Department of Physics and Astronomy, JHU, Baltimore, MD 21218, USA}
\author{Kadin Worthen}\affiliation{Center for Astrophysical Sciences, The William H. Miller III Department of Physics and Astronomy, JHU, Baltimore, MD 21218, USA}
\author{Kielan Hoch}\affiliation{Space Telescope Science Institute, 3700 San Martin Drive, Baltimore, MD 21218, USA}
\author{Christine Chen}\affiliation{Space Telescope Science Institute, 3700 San Martin Drive, Baltimore, MD 21218, USA}
\author{Stephanie Sallum}\affiliation{Department of Physics and Astronomy, University of California, Irvine, Irvine, CA, USA}
\author[0000-0003-2259-3911]{Shrishmoy Ray}\affiliation{University of Exeter, Astrophysics Group, Physics Building, Stocker Road, Exeter, EX4 4QL, UK.}
\author[0000-0002-2805-7338]{Karl Stapelfeldt}\affiliation{Jet Propulsion Laboratory, California Institute of Technology, M/S 321-100, 4800 Oak Grove Drive, Pasadena, CA 91109, USA}
\author{Yifan Zhou}\affiliation{Department of Astronomy, The University of Texas at Austin, 2515 Speedway Boulevard Stop C1400, Austin, TX 78712, USA}
\author{Michael Meyer}\affiliation{Department of Astronomy, University of Michigan, Ann Arbor, MI 48109, USA}
\author{Mickael Bonnefoy}\affiliation{Universit\`e Grenoble Alpes / CNRS, Institut de Plan\`etologie et d’Astrophysique de Grenoble, 38000 Grenoble, France}
\author{Thayne Currie}\affiliation{University of Texas, San Antonio}
\author{Camilla Danielski}\affiliation{Instituto de Astrofísica de Andalucía, CSIC, Glorieta de la Astronomía, 18008, Granada, Spain}\affiliation{AIM, CEA, CNRS, Universit\`e Paris-Saclay, Universit\`e Paris Diderot, Sorbonne Paris Cit\`e, F-91191 Gif-sur-Yvette, France}
\author[0000-0003-0593-1560]{Elisabeth C.~Matthews}\affiliation{Observatoire Astronomique de l’Universit\'e de Gen\`eve, 51 Ch.~Pegasi, 1290 Versoix, Switzerland}
\author[0000-0003-1251-4124]{Anand Sivaramakrishnan}\affiliation{Space Telescope Science Institute, 3700 San Martin Drive, Baltimore, MD 21218, USA}
\author[0000-0001-7864-308X]{Rachel A. Cooper} \affiliation{Space Telescope Science Institute, 3700 San Martin Drive, Baltimore, MD 21218, USA}
\author[0000-0002-1536-7193]{Deepashri Thatte}
\affiliation{Space Telescope Science Institute, 3700 San Martin Drive, Baltimore, MD 21218, USA}
\author[0000-0003-0454-3718]{Jordan Stone}\affiliation{US Naval Research Laboratory, Remote Sensing Division, 4555 Overlook Ave SW, Washington, DC 20375}

\author{Malavika Vasist}\affiliation{Space Sciences, Technologies, and Astrophysics Research (STAR) Institute, University of Li\`ege, B-4000 Li\`ege Belgium}



\begin{abstract}
We present a set of recommended best practices for JWST data collection for members of the community focussed on the direct imaging and spectroscopy of exoplanetary systems. These findings and recommendations are based on the early analysis of the JWST Early Release Science Program 1386, ``High-Contrast Imaging of Exoplanets and Exoplanetary Systems with JWST.''  Our goal is for this information to be useful for observers in preparation of JWST proposals for Cycle 2 and beyond.  In addition to compiling a set of best practices from our ERS program, in a few cases we also draw on the expertise gained within the instrument commissioning programs, as well as include a handful of data processing best practices. We anticipate that this document will be regularly updated and resubmitted to the \textit{arXiv} to ensure that we have distributed our knowledge of best-practices for data collection as widely and efficiently as possible.  
\end{abstract}



\section{Introduction} \label{sec:intro}
With a combination of unprecedented sensitivity and wavelength coverage \textit{JWST} has already demonstrated its potential for carrying out potentially transformative science related to the detection and characterization of exoplanetary systems.  The exquisite wave front stability, as well as the capability to observe exoplanet atmospheres near the peak of their thermal emission at 3-5\,$\mu$m and beyond, means that JWST will have extraordinary power going forward for characterizing wide-separation exoplanets through direct imaging. The precise calibration of the observatory Point Spread Function (PSF) that is afforded by this spacecraft stability \cite[e.g.,][]{ppv18}, combined with sensitivity in the near- and mid-infrared that is in some cases hundreds of times greater than that for ground-based instruments means that JWST will likely be sensitive to an entirely new class of sub-Jupiter mass planets at wide orbital separations \citep[e.g.,][]{chb21}.  This stability and wavelength coverage, combined with wide fields of view in JWST imagers, will also allow resolved imaging of circumstellar disks at wavelengths largely out-of-reach to ground-based observatories.   The broad wavelength coverage will also allow the JWST spectrographs within NIRSpec and MIRI to cover multiple spectroscopic features of exoplanet atmospheres, giving precise measurements of atmospheric compositions and atmospheric chemistries.  

In 2017, the Space Telescope Science Institute (STScI) awarded roughly 500 hours of Director's Discretionary Time to 13 community-driven Early Release Science (ERS) programs with the goals of: 1) testing the observatory in the modes expected to be commonly used by that community; and 2) to make clear recommendations to the community on best-practices to be used in future cycles as well as distribute a set of Science Enabling Products (SEPs).  Our program ``High-Contrast Imaging of Exoplanets and Exoplanetary Systems with JWST'' \citep{hcr22}, was ultimately awarded 55 hours of DDT time to utilize all four JWST instruments, and assess the performance of the observatory in these representative modes. The first science products from this program have already been presented in two initial publications.   \cite{chk22} showcases the first-ever direct images of an exoplanet with JWST,  as well as the first images of an exoplanet at wavelengths longer than 5\,$\mu$m.  \cite{mbp22} shows the first direct spectrum of a planetary mass comapanion with JWST, and the first spectrum of such an object covering its full luminous range from 1 to $\sim$20\,$\mu$m, clearly showing evidence for disequilibrium chemistry and the first definitive detection of silicate clouds in an atmosphere of a planetary mass companion.  

In this brief document our aim is to compile a set of data collection best practices for members of the community focussed on the direct detection and spectroscopy of exoplanetary systems. Our goal is that this information will be useful for preparation of proposals for Cycle 2 and beyond. In a few instances we also make recommendations for data processing best practices.  In addition to compiling a set of best practices from our ERS program, in a few cases we also draw on the expertise gained within the JWST commissioning programs \citep[e.g.,][]{glk22, kgc22, kcv22, stl22}. We anticipate that this document will be updated and resubmitted to the \textit{arXiv}, perhaps multiple times, to ensure that we have distributed our knowledge of best-practices for data collection, as efficiently as possible.  In \S\ref{sec:exoplanet_coronagraphy} and \ref{sec:diskcoronagraphy}, we outline our recommendations for best practices for carrying out NIRCam and MIRI coronagraphy of point sources and extended (e.g., disk) structures, respectively.  In \S\ref{sec:spectroscopy} we outline our recommendations for spectroscopy of point sources using NIRSpec and the MIRI Medium Resolution Spectrgraph (MRS), and in \S\ref{sec:ami} we highlight some best practices for employing Aperture Masking Interferometry (AMI) with the NIRISS instrument.

\section{Recommendations for Point Source Coronagraphy}\label{sec:exoplanet_coronagraphy}

In this section, we outline our recommendations for the best practices for carrying out \textit{point source} coronagraphy (e.g., exoplanets and substellar companions to bright stars) with both NIRCam and MIRI, as well as some recommendations that can apply to both instruments. 

\subsection{NIRCam Coronagraphy}

\begin{itemize}
    \item \textbf{Dual-band Imaging:} Unlike the first observation cycle, dual-band imaging will be enabled for NIRCam Coronagraphic imaging for Cycle 2. This will allow for simultaneous observations in the Short Wavelength (SW) and Long Wavelength (LW) channels and the filters in each. Only one coronagraphic mask can be used for both channels.   
 
    \item \textbf{Inner Working Angle:} Because the NIRCam round coronagraphic masks  have a smooth transmission function, the ``inner working angle'' (IWA) of the coronagraph is stated to be where the coronagraph has 50\% transmission. However, as discussed in \cite{kgc22} and \cite{glk22}, the stability of the JWST wavefront means that objects interior to this IWA can be detected at high confidence, depending on the particulars of the target.
    
    \item \textbf{Choice of Coronagraphic Mask:} The MASK335R coronagraphic mask currently has the best Target Acquisition (TA) performance \citep[e.g.,][]{glk22}, which translates directly into somewhat superior coronagraphic suppression.  If given the choice between the MASK335R and MASK430R, proposers should choose the former. 
    
    \item \textbf{Small Grid Dithers (SGD):} The use of SGDs is still highly recommended to compensate for TA errors, at least when searching for faint sources within $\sim$2 arcsec.
\end{itemize}

\subsection{MIRI Coronagraphy}
\begin{itemize}
    \item \textbf{Stray-light Artifacts:} Data obtained during commissioning and the ERS period revealed that MIRI coronagraphic images are significantly impacted by scattered light, producing bright linear features at the boundaries of the Four Quadrant Phase Mask (4QPM).  This scattered stray-light, sometimes referred to informally as the ``glow stick'' feature and shown in Figure 2 of \cite{chk22}, significantly impacts all MIRI coronagraphic observations. To mitigate this very serious systematic, dedicated background observations are required for Cycle 2, with exposure times matching those of the target, as well as any reference observations. It is also highly recommended that observers use the automatic extra observation feature in APT to obtain two backgrounds, one each in opposing quadrants. This enables a basic dither and mitigates against having a background astrophysical source fall on the glow stick that could severely hamper the glow-stick correction.
    
    \item \textbf{Target Acquisition:} TA is currently only available in Quadrant 1 of each of the MIRI coronagraphs. The MIRI Instrument Team is commissioning TAs for Quadrant 4 during Cycle 1 observations. Observers who may need TA in this quadrant should contact the JWST Help Desk.
    
    \item \textbf{Neutral Density Filters for Target Acquisitions:} TA is currently only available to be used with the Neutral Density Filter (FND). The other three TA filters (F560W, F1000W and F1500W) are being commissioned during Cycle 1, starting with the F560W. Observers needing one of these filters for TA should contact the JWST Help Desk.
    
    \item \textbf{Special Considerations for Cycle 2:} Additional information on MIRI coronagraphic imaging, including some considerations that should be taken into account for Cycle 2 can be found in the on the \href{https://jwst-docs.stsci.edu/jwst-mid-infrared-instrument/miri-operations/miri-target-acquisition/miri-coronagraphic-imaging-target-acquisition}{JWST User Documentation} (``JDox.'').
\end{itemize}

\subsection{Considerations for Both NIRCam and MIRI Point Source Coronagraphy}\label{sec:NIRCam_MIRI_coronagraphy}
\begin{itemize}
    \item\textbf{Reference Star Vetting:} We urge future observers to exercise caution in their selection of reference stars for coronagraphy using all possible archival information (e.g., previous AO imaging, Gaia RUWE statistic), as well as additional  ground-based supporting observations, to ensure that reference stars are indeed single stars. This applies, of course, also to coronagraphy of circumstellar disks as well discussed in \S\ref{sec:diskcoronagraphy}.  Ground-based observations such as speckle-imaging, AO Imaging, or Aperture Masking Interferometry (AMI) observations can be highly effective for eliminating reference stars that might have contaminating additional objects within $\sim$0.5 - 1.0$^{\prime\prime}$. 
  
    \item\textbf{ADI+RDI versus RDI Alone:} As discussed in \cite{chk22}, the combined ADI+RDI KLIP PSF subtraction did not provide significantly improved contrast performance over a solely RDI-based subtraction for the ERS datasets. As a result, users may be able to improve observational efficiency by opting for observations that facilitate an RDI-based subtraction alone, especially for targets within $\sim$500\,mas to $\sim$1 arcsec. However, at large angular separations from the star, users may find that using ADI alone will be adequate to remove any residual scattered starlight.
    
    \item\textbf{Additional Tools:} The default ETC offers a means to simulate coronagraphic performance, but only for simplistic RDI subtractions with no small-grid dithers. Users should consider investigating additional tools such as \href{https://aarynncarter.com/PanCAKE/index.html}{PanCAKE}, \href{https://github.com/JarronL/pynrc/tree/develop}{pyNRC}, MIRISim, and NIRCCoS to produce more accurate estimates of the achievable performance across a wider range of observational strategies. 
    \item Detailed information on a variety of observational considerations and recommendations can be found in the \href{https://jwst-docs.stsci.edu}{JWST Documentation}.  The specific links might be helpful related to 
\begin{itemize}
    \item \href{https://jwst-docs.stsci.edu/jwst-near-infrared-camera/nircam-observing-strategies/nircam-coronagraphic-imaging-recommended-strategies}{NIRCam Coronagraphic Imaging}, 
    \item \href{https://jwst-docs.stsci.edu/jwst-mid-infrared-instrument/miri-observing-strategies/miri-coronagraphic-recommended-strategies}{MIRI Coronagraphic Imaging}, as well as 
    \item \href{https://jwst-docs.stsci.edu/jwst-mid-infrared-instrument/miri-observing-strategies/miri-cross-mode-recommended-strategies}{MIRI Cross-Mode Strategies}
    \end{itemize}
\end{itemize}


\section{Best Practices for Coronagraphy of Circumstellar Disks}\label{sec:diskcoronagraphy}
The ERS program includes coronagraphic observations with both NIRCam and MIRI of a benchmark debris disk system, HD\,141569A \citep[e.g.,][]{wrb00}, which has been well-studied from both ground-based \citep[e.g.,][]{blr15, bmd20, sbb21} as well as space-based imaging \citep{cka03, wbs99}.  This target is comprised of at least three concentric disk rings, and is therefore also an ideal target to test the sensitivity of JWST to extended structures as a function of angular distance from the host star, as well as study the impact of various post-processing algorithms on extended structures.  

While the NIRCam coronagraphic data for this object have yet to be gathered at the time of this writing, the MIRI data were successfully obtained on 2022 August 02. The processed and combined data show an image in all three filters dominated by the thermal emission of the inner ring ($r{\sim}$0.4$^{\prime\prime}$). The emission of the outer rings ($r{\sim}$1.7$^{\prime\prime}$ and $r{\sim}$3.2$^{\prime\prime}$) are hidden by the complex response of the bright inner disk, which extends out in a diffuse halo out to ${\sim}6^{\prime\prime}$. This response is difficult to simulate and subtract off from the images because the ring lies near the inner working angle of the 4QPM mask and is neither separated from, or  fully masked by, the coronagraph. Thus, the poor knowledge of the spatial distribution of the dust emission makes it difficult to disentangle the complex instrument response from possible asymmetries in the disk surface brightness.

\textbf{Reference Star Vetting:} As emphasized in \S\ref{sec:NIRCam_MIRI_coronagraphy}, as a blanket statement covering both NIRCam and MIRI disk coronagraphy, we urge future observers to carefully select reference stars for MIRI coronagraphy taking advantage of all possible archival information, as well as potential ground-based supporting high-resolution observations (e.g., speckle, AO, AMI) to ensure that reference stars are indeed single stars. 

\subsection{NIRCam coronagraphy Best-Practices for Disk Observations}
    \textbf{Multiple Systems:} Observers should exercise caution in observing targets that are members of multiple systems, even if the angular distance of any additional companions initially seems too large to be significant, due to the risk of unexpected scattered light or diffraction effects from neighboring stars interfering with the primary science observations.  
    HD 141569A (A2V) has two M-star companions, closely positioned on the sky, and both at a separation of ${\sim}8^{\prime\prime}$.  Since the HD141569A primary star is much brighter than these M-dwarf companions, these two objects have not typically posed a problem for disk observations in previous high contrast imaging with HST/STIS and ground-based instruments. However, at the longer wavelengths probed by JWST these objects are much more comparable in brightness. Simulations of our upcoming NIRCam observations indicate that the diffraction patterns of these stars at the location of the disk are of comparable (or greater) brightness to the disk. As a result, we have added offset observations of our reference star to our program at a position such that we plan to use it to subtract off these diffraction patterns. This strategy will be tested once the observations are obtained later in 2023.     

\subsection{MIRI Coronagraphy Best-Practices for Circumstellar Disk Observations}
\begin{itemize}
    \item \textbf{Bright Dust Emission Close to Central Star:} Some systems may exhibit bright emission of circumstellar dust close to the star (with an angular extent still inside the inner working angle of the coronagraphs).  
    As was seen in our ERS program, dust that is resolved (or partially resolved) inside of the 4QPM inner working angle may deviate sufficiently from a point-like structure, introducing significant diffraction patterns such that the coronagraphic suppression may be compromised, and thereby limiting sensitivity. This has so far proven difficult to forward model and subtract, but is an active area of research in the ERS team. 

    \item\textbf{Ground-based Supporting Observations:} One potential protection against this would be to obtain supporting ground-based observations (e.g., VISIR at VLT) to test for such resolved emission from inner dust structures. 

    \item\textbf{JWST Direct Images:} If there is \textit{clear} evidence (perhaps from supporting ground-based observations) for a bright, spatially-extended inner disk, then \textit{direct (i.e. non-coronagraphic) images} using MIRI may prove advantageous to help forward-model the disk signal in coronagraphic modes. Indeed, with appropriate forward modeling a good inner disk model may enable subtraction of the extra diffraction introduced.   Our team is currently exploring this. 
    
    \item\textbf{A Warning About Saturation:} Bright targets may saturate normal direct imaging mode MIRI filters, which may complicate the direct imaging approach discussed above. One potential option is to use coronagraphic TA images with the FND filter (not available to direct imaging observations). However the wavelength range is extremely broad and the data requires significant forward modeling to interpret. 

\end{itemize}


\section{NIRSpec \& MIRI MRS Spectroscopy of Point Sources}\label{sec:spectroscopy}

Below we highlight some aspects of data collection, as well as a few for data processing,  that have arisen as part of this ERS program.  More details can be found in \cite{mbp22}, and the fully processed spectrum of the wide-separation brown dwarf VHS\,1256b, a relatively bright point source, can be delivered upon request.  

\subsection{Best Practices for Point Source Spectroscopy Data Collection using the NIRSpec IFS}

The ERS observations of VHS\,1256b using the NIRSpec IFS are photon noise limited. Currently, any limiting factor for the NIRSpec IFS data quality for this source is the data processing carried out by the JWST pipeline.  This is expected to be improved by further Cycle 1 calibration observations and improvements to the IFS cube building algorithm.

\subsubsection{NIRSpec IFS Best Practices for Data Collection}

\begin{itemize}
\item \textbf{NIRSpec Dithering:} The option employing the four-point nod was used to ideally improve spatial and spectral sampling which is consistent with the recommendations contained within the \href{https://jwst-docs.stsci.edu/jwst-near-infrared-spectrograph/nirspec-observing-strategies/nirspec-dithering-recommended-strategies}{NIRSpec Dithering decision flowchart}.  
\end{itemize}

\subsubsection{NIRSpec IFS Post-Processing Considerations for Point Source Spectroscopy}

\begin{itemize}
\item\textbf{Hot pixels and other sources:} Currently the data quality flags of the JWST Pipeline do not adequately flag and remove hot pixels from detector images. These extremely high values are propagated through the image cubes where spectra are extracted. In the work presented in \cite{mbp22} this was addressed by reducing dithers individually and taking a weighted mean to reduce the likelihood of hot pixels influencing the final spectra.

\item\textbf{Flux Calibration and Spectral Response:} The flux calibration and spectral response correction produced by the default JWST pipeline may not give reliable results even for standard stars. When flux calibrating and stitching together observation modes of your target use a correction factor based on standard targets (e.g., A-stars) with a known spectral response.

\item\textbf{Oscillations within Spectra} - An oscillating structure is present in spectra extracted from spectral cubes. Some of these oscillations result from the cube building algorithm itself or potentially arise from the undersampled PSF of NIRSpec. These features were not addressed in \cite{mbp22} but need more work to understand why they are created in the Stage 3 cube building step of the pipeline.

\item\textbf{Spectral Extraction:} - The extraction step of the JWST pipeline for NIRSpec does not let the user define extraction widths and no aperture correction exists for NIRSpec at the moment. 
\end{itemize}

\subsection{MIRI MRS Spectroscopy of Point Sources}
The ERS MIRI observations for VHS\,1256b are background limited compared to the NIRSpec observations. A significant source of noise in the observation of faint, isolated point sources is prominent detector noise (vertical striping on the detector), as well as cosmic ray showers. Mitigations steps for these are described below: 

\begin{itemize}
    \item\textbf{Background observations:} Background observations were not used directly in the processing carried out by the ERS team, since the contamination of cosmic ray showers introduced artefacts in the science images. The background was instead estimated with a reference aperture off source in the reconstructed data cubes. An annulus background estimation would also have been appropriate.

    \item\textbf{Vertical Detector Striping:} The vertical detector stripes are in the same direction of the dispersion, therefore very difficult to disentangle from the spectrum. The background observations can be used to estimate this artifact and remove it from the science frames (this is relevant mainly for the IFUSHORT detector). Background observations can also detect remaining hot/warm pixels that are not flagged by the pipeline.

    \item\textbf{Cosmic Rays:} Cosmic ray showers appear as diffuse structures at similar levels as the point source. The pipeline is not able to flag these during the ramp fitting step. Therefore the best solution at the moment is to rely on the information of the different dithers (4 being ideal) to average out the effect of cosmic rays. 
\end{itemize}

Thus, our summary recommendation for MIRI MRS observations of point sources is to 1) obtain background observations, assuming time allows for this; 2) hope for the absence of high-intensity cosmic ray showers; and 3) use the four-point dither pattern.  All other calibration steps are handled very well by the STScI pipeline.  For any remaining features in the spectra, residual fringes may be the cause.


\section{Best Practices for Aperture Masking Interferometry with NIRISS}\label{sec:ami}
Observations of the exoplanet host star HIP65426 using Aperture Masking Interferometry \citep[e.g.,][]{tmd00,i13,ss19} with the NIRISS instrument were carried out within this program, as described in \cite{hcr22}.  Observations were obtained on 2022 July 30. Below we highlight some findings and recommendations from our ERS program, but much more information on NIRISS AMI and commissioning can be found in \cite{stl22}.  

\begin{itemize}

\item \textbf{Achievable Contrast (Compared to Coronagraphy):} NIRISS operating in the AMI mode outperforms NIRCam coronagraphy within $\sim$300-500\,mas, with a contrast curve that is essentially flat between this region and $\sim$100\,mas. 

\item \textbf{Achievable Contrast (Compared to Theoretical Photon Noise Limits):} Initial analysis of both ERS and commissioning data has shown that the achieved contrast with AMI is at least one magnitude poorer than the theoretical prediction based on photon noise that can be found in the \href{https://jwst-docs.stsci.edu/jwst-near-infrared-imager-and-slitless-spectrograph/niriss-observing-strategies/niriss-ami-recommended-strategies#NIRISSAMIRecommendedStrategies-Exposuredepthestimationforbinarypointsource}{AMI Recommended Strategies} in JDox.  

\item \textbf{Reference PSF Brightnesses and Detector Settings:} Observations of the target and any reference calibration stars should have a similar number of groups and integrations so that the photon noise limit is reached in the analysis of the target. In other words, reference and science should have equal brightness in the observed filter, with the goal of achieving a similar level of systematic effects such as charge migration.  

\item \textbf{Reference PSF Spectral Types:} Initial analysis of ERS observations of HIP65426, as well as commissioning observations \citep{kcv22, stl22} indicate that matching the number of groups and integrations has a greater impact on the overall calibrations, than matching the spectral types between target and calibrator.

\item \textbf{Reference PSF Vetting:} In general it is good practice to select well-vetted reference PSF calibrators when possible, to avoid potential contamination by binary calibrators. When this is not possible, selecting multiple calibrator stars can help to mitigate this risk.

\item \textbf{Optimizing Target Placement:} Target acquisition during AMI commissioning showed highly repeatable filter-dependent systematic placement errors of up to a third of a pixel.  Improved centering within a pixel may well reduce AMI sensitivity to flat-fielding error, jitter, and other effects.  An offset for an observation should be specified (as a Special Requirement) in AMI's APT template to place targets close to the pixel center.  Given the observation's filters, offsets can be chosen after an inspection of Figure 5 in \citet{stl22}. 

\item \textbf{Frame Rejection:} Commissioning and ERS data analysis often  excluded 10\% of integrations (stringent data cleaning rejects 30\%).  A tool to identify and ignore or repair affected parts of the 2-dimensional images in cal\_ints data cubes is under development.
\end{itemize}

\bibliography{MasterBiblio_Sasha}
\bibliographystyle{aasjournal}



\end{document}